\documentclass[12pt]{article}
\usepackage[utf8]{inputenc}
\usepackage{amsmath}
\numberwithin{equation}{section}
\usepackage{amsfonts}
\usepackage{amssymb}
\usepackage{mathrsfs}
\usepackage{slashed}
\usepackage{indentfirst}
\usepackage{graphics}
\usepackage[framemethod=TikZ]{mdframed}
\mdfdefinestyle{MyFrame}{%
    linecolor=blue,
    outerlinewidth=2pt,
    roundcorner=20pt,
    innertopmargin=\baselineskip,
    innerbottommargin=\baselineskip,
    innerrightmargin=20pt,
    innerleftmargin=20pt,
    backgroundcolor=white!50!white }
      
\begin{document}
\title{Unified prescription for the generation of electoweak and gravitational gauge field Lagrangian on a principal fiber bundle}
\author{Theo Verwimp \footnote{Former affiliated with: Physics Department; U.I.A., Universiteit Antwerpen Belgium. On retirement from ENGIE Laborelec, Belgium.}\\
e-mail: theo.verwimp@telenet.be}
\renewcommand{\today}{October 14, 2021}
\maketitle
The Glashow-Weinberg-Salam gauge field Lagrangian for
electroweak theory and the Townsend-Zardecki action for gravitation are obtained from the
same type of Yang-Mills Weil form on a principal fiber bundle over space-time, with
symmetry group U(2) and SO(2,3), respectively. The unified geometrical approach given here
shows that fiber bundle reduction and symmetry breaking are essential not only in electroweak theory but also in the SO(2,3) gauge theory for gravitation. In fact, the process of symmetry breaking in electroweak theory and the soldering of the anti-de Sitter bundle, essential in the interpretation of SO(2,3) gauge theory as a theory for gravitation, are corresponding geometrical concepts. 

\section{Introduction}
It has been long known that from the unifying point of view of fiber bundles, the kinematics of gauge theories and general relativity are the same.$^{[1]}$ However, the gauge scheme itself fails to determine directly the dynamics of the theory, i.e., a form of Lagrangian. Euler-Lagrange equations are obtained after a Lagrangian has been defined. For example, in a Yang-Mills theory with a Lie group $G$, the gauge field Lagrangian is defined as
\begin{equation}
I_{AB}F^{A}_{\mu\nu}F^{B\mu\nu},
\end{equation}
where $F^{A}_{\mu\nu}$ are the components of the curvature tensor $F^{\mu\nu}$ of the gauge field relative to a basis of $\mathscr{G}$ , the Lie algebra of $G$, and $I_{AB}$ is the metric tensor of the bilinear Killing form of $G$, $A, B : 1,...,k$, $k$ the dimension of $\mathscr{G}$.$^{[2]}$ Within the fiber bundle formulation of gauge theories it is, however, more natural to define a Lagrangian form on the principal fiber bundle $P(M,G)$ over space-time $M$.

A most general gauge invariant Lagrangian form on $P$ that projects uniquely to the base manifold can be defined using Weil forms$^{[3]}$ constructed from Weil polynomials$^{[4]}$ and $\mathscr{G}$-valued tensorial two-forms on $P$. Weil polynomials were used by Kakazu and Matsumoto in the reconstruction of Einstein gravity (with torsion) and four-dimensional $N=1$ supergravity on a principal fiber bundle (PFB).$^{[5]}$ It was also possible to reconstruct the Lovelock Lagrangian$^{[6]}$ for higher dimensional gravity on the bundle of orthonormal frames over a space-time of arbitrary dimension D by using Weil polynomials on the Lie algebra $so(1,D-1)$.$^{[3]}$

In this paper, we will emphasize the complete analogy between the derivation of the Glashow-Weinberg-Salam gauge field Lagrangian for electroweak theory and the Townsend-Zardecki type action for gravitation$^{[7]}$ from the same type of Weil form on a PFB with symmetry group
$U(2)$ and $SO (2,3)$, respectively. Besides the Einstein action
and a cosmological term, the Townsend-Zardecki action also contains curvature-squared and torsion-squared terms. However, for zero torsion and zero effective cosmological constant, the theory reduces essentially to Einstein's theory in vacuum.$^{[7]}$ Both Townsend and Zardecki derive their Lagrangian in the context of a $SO(1,4)$ gauge theory. Here, we will consider as gauge group for gravitation the anti-de Sitter group $SO(2,3)$.[8]

The rest of the paper is organized as follows. In Sec.2
we describe the reduction of a principal fiber bundle $P$ to a
subbundle $Q$, the splitting of a connection one-form on $P$ into
a reduced connection on $Q$ and a tensorial one-form, and give the expressions for the reduced curvature on $Q$. The occurrence of reduction is closely related with the process of symmetry breaking. In the electroweak theory the remaining symmetry is that of the charge conservation group $U_{c}(1)$, while for gravitation we consider symmetry breaking from $SO(2,3)$ to $SO(1,3)$. In Sec.3 we outline the construction of a gauge invariant Lagrangian form on the $U(2)$ and the $SO(2,3)$ principal fiber bundle, respectively, and determine the Weil polynomials needed in the construction of these Lagrangian forms. With the theory thus developed it is then straightforward to determine the electroweak and gravitational gauge field Lagrangian in Secs.4 and 5. Although the derivation of the gravitational gauge field Lagrangian is, except for the gauge group, not essentially different from that given by Zardecki,$^{[7]}$ our unified geometrical approach, however, will make it possible to identify corresponding fields and concepts in the electroweak and gravitational gauge theories.

\section{Fiber bundle reduction}
Let $P(M,G)$ be a PFB with base space $M$ and structure group $G$. If $\rho :G\rightarrow GL(V)$ is a representation of $G$ in a finite dimensional vector space $V$, then a pseudotensorial form of degree $r$
 on $P$ of type $(\rho,V)$ is an $r$-form $\psi:P\rightarrow V$, which is equivariant with respect to the action of $G$, i.e.,
 \begin{equation}
 R^{\ast}_{g}\psi =\rho(g^{-1})\cdot\psi,
 \end{equation}
where $R_{g}$ is the right action of $g\in G$ on $P$. Such a form $\psi$ is
called a tensorial form if $\psi(X_{1},...,X_{r})=0$ whenever at least  one of the tangent vectors $X_{i}$ of $P$ is tangent to a fiber.$^{[4]}$ In particular, a tensorial zero-form of type $(\rho,V)$ is a function $\psi:P\rightarrow V$ such that
\begin{equation}
\psi(ug)=\rho(g^{-1})\cdot\psi(u), \quad \, u\in P, \, g\in G.
\end{equation}

Suppose that $H$ is a closed subgroup of $G$. On the coset
space $G/H$, the Lie group $G$ acts as a transitive Lie transformation group, i.e., $G/H$ is a homogeneous manifold of $G$. If $G/H\subset V$ the vector space on which $G$ acts through $\rho$, then the coset space $G/H$ can be thought of as the orbit space $V_{0}=\rho(G)\cdot v_{0}$ with $v_{0}\in G/H$ an $H$-fixed point in $V$. Now, let $E(M,G/H,G,P)$ be the vector bundle associated with $P$ and with standard fiber $G/H$. Then one has the following result:$^{[4]}$ the bundle $P(M,G)$ has a reduction to a subbundle $Q(M,H)$ iff $E$ admits a global section. Moreover, there is a one-to-one correspondence between cross sections $\sigma:M\rightarrow E$ and equivariant mappings $\psi:P\rightarrow G/H$. Therefore, a necessary and sufficient condition for the occurrence of the reduction is the
existence on $P$ of a tensorial zero-form $\psi$ of type $(\rho,V)$ with
range $\psi(P)$ in the orbit space $G/H$ (see also Ref. 10). Thereby $Q =\psi^{-1}(v_{0})$, and Fulp and Norris in Ref. 11 refer to $\psi$ as
a symmetry breaking Higgs field.

If the symmetry $G$ is a gauge symmetry associated with a
Lagrangian field theoretical model, then the original symmetry $G$ is said to be broken spontaneously by the Higgs mechanism iff the self-interaction potential $V(\Psi)$ of a multiplet $\Psi$ of (possibly complex) scalar fields, which are described by a global section of a vector bundle $W$ associated with $P$ with fiber $V\supset G/H$, is minimized by the manifold $V_{0}=G/H$. In that case, the global section $\Psi_{0}(x)=v_{0}$, is called the vacuum or ground state and the homogeneous space $V_{0}$, obtained by the action of $\rho(G)$ on $v_{0}$, the vacuum manifold.

Let $\gamma:Q\rightarrow P$ be the imbedding of the reduced bundle $Q(M,H)$ into its extension $P(M,G)$, and set
\begin{equation}
\mathscr{G}=\gamma_{\ast}(\mathscr{H})\oplus\mathscr{T},
\end{equation}
where $\mathscr{H}$ is the Lie algebra of the subgroup $H$ and $\mathscr{T}$ a vector subspace of $\mathscr{G}$, the Lie algebra of $G$. If $\tilde{\mu}$ is a connection form on $P$, then the restriction $\mu$ of $\tilde{\mu}$ to $Q$, i.e., $\mu=\gamma^{\ast}\tilde{\mu}$, splits naturally into
\begin{equation}
\mu = \omega +\phi,
\end{equation}
where $\omega$ is $\mathscr{H}$ valued and $\phi$ is $\mathscr{T}$ valued. We then have the theorem [12] that if the complement $\mathscr{T}$ of $\mathscr{H}$ in $\mathscr{G}$ is Ad invariant by $H$ or equivalently $\left[ \gamma_{\ast}\mathscr{H},\mathscr{T}\right]\subset\mathscr{T} $ if $H$ is connected, then the restriction $\omega$ of the $\mathscr{H}$ component of the connection one-form $\tilde{\mu}$ on $P$ is a connection one-form on $Q$ while $\phi$ is a tensorial one-form of type  $(Ad H,\mathscr{T)}$ on $Q$. Here, $Ad$ always denotes the adjoint representation of the symmetry group on its Lie algebra.

The curvature two-form $\tilde{\Delta}$ on $P$ is a tensorial two-form of type $(Ad G,\mathscr{G})$ defined as
\begin{equation}
\tilde{\Delta}=D\tilde{\mu}=d\tilde{\mu}+\frac{1}{2}\left[\tilde{\mu},\tilde{\mu}\right],  
\end{equation}
with $\left[,\right]$ denoting the exterior product of forms with values in a Lie algebra. The reduction $\Delta=\gamma^{\ast}\tilde{\Delta}$ to $Q$ of the curvature $\tilde{\Delta}$ calculated from $\tilde{\mu}$ on $P$ can be written$^{[5]}$
\begin{equation}
\Delta=\Omega +\Phi +\Sigma,
\end{equation}
where
\begin{align}
&\Omega=d\omega+\frac{1}{2}\left[ \omega,\omega\right], \\
&\Phi=d\phi+\left[ \omega,\phi\right], \\
&\Sigma=\frac{1}{2}\left[ \phi,\phi\right].
\end{align}
From the Bianchi identity on $P$
\begin{equation}
d\tilde{\Delta}+[\tilde{\mu},\tilde{\Delta}]=0,  
\end{equation}
we find after reduction to the subbundle $Q$
\begin{align}
&D\Omega=d\Omega+\left[ \omega,\Omega\right]=0,\\
&D\Phi=d\Phi+\left[ \omega,\Phi\right]=\left[ \Omega,\phi\right].
\end{align}

\section{A gauge invariant Lagrangian form}
In general, $Ad(G)$-invariant Weil polynomials$^{[4,5,13]}$ of degree $m$ on the Lie algebra $\mathscr{G}$ are defined as multininear symmetric real-valued functions $L_{m}$ such that
\begin{equation}
L_{m}(g\cdot T_{1},...,g\cdot T_{m})=L_{m}(T_{1},..., T_{m}),
\end{equation}
for all $g\in G$, $T_{i}\in\mathscr{G} $ and where $g\cdot T_{i}=gT_{i}g^{-1}$. The space of all Weil polynomials of degree $m$ is denoted by $S^{m}_{G}(\mathscr{G})$.

If $\tilde{\psi}_{1},...,\tilde{\psi}_{m}$ are $\mathscr{G}$-valued two-forms on $P$, then we define for each $L_{m}\in S^{m}_{G}(\mathscr{G}) $ a real valued Weil form $L_{m}(\tilde{\psi}_{1},...,\tilde{\psi}_{m})$ on $P$ of degree $2m=D$ the dimension of $M$, by
\begin{multline}
L_{m}(\tilde{\psi}_{1},...,\tilde{\psi}_{m})(X_{1},...,X_{D})=\\\left( \frac{1}{2}\right)^{m}\sum_{\sigma\in S_{D}}\epsilon(\sigma)L_{m}\left( \tilde{\psi}_{1}(X_{\sigma(1)},X_{\sigma(2)}),...,\tilde{\psi}_{m}(X_{\sigma(D-1)},X_{\sigma(D)})\right), 
\end{multline}
for $X_{1},...,X_{D}\in T_{u}(P)$ (tangent space of $P$ at $u$), where the summation is taken over all permutations $\sigma$ of $1,...,D)$ and $\epsilon (\sigma)$ is the sign of the permutation. If $\left\lbrace T_{A}\right\rbrace $ is a basis for $\mathscr{G}$ such that $\tilde{\psi}_{i}=\tilde{\psi}_{i}^{A}T_{A}$ [$\tilde{\psi}_{i}^{A}\in\Lambda^{2}(P,\mathbb{R})$ the space of real-valued two-forms on $P$], then one obtains from the multilinearity of $L_{m}$ and the definition of the wedge product that
\begin{equation}
L_{m}(\tilde{\psi}_{1},...,\tilde{\psi}_{m})=L_{m}(T_{A_{1}},...,T_{A_{m}})\tilde{\psi}_{1}^{A_{1}}\wedge\cdot\cdot\cdot\wedge\tilde{\psi}_{m}^{A_{m}}.
\end{equation}
If the restrictions $\psi_{i}=\gamma^{\ast}\tilde{\psi}_{i}$ are tensorial forms on $Q$, then the restriction of the Weil form itself, i.e.,
\begin{equation}
L_{m}(\psi_{i},...,\psi_{m})=\gamma^{\ast}L_{m}(\tilde{\psi}_{1},...,\tilde{\psi}_{m})
\end{equation}
will project to a unique $D$-form $\mathscr{L}$ on $M$ such that (Ref.4, Chap.XII)
\begin{equation}
L_{m}(\psi_{i},...,\psi_{m})=\pi^{\ast}\mathscr{L},
\end{equation}
where $\pi$ is the projection from the subbundle $Q$ on the base manifold $M$.

To determine the dynamics of a gauge theory, a gauge field Lagrangian must be defined. The gauge invariant Lagrangian form on $P(M,G)$, $M$ the space-time manifold, given by
\begin{equation}
L_{2}(\tilde{\Delta},\ast\tilde{\Delta}),
\end{equation}
where $\tilde{\Delta}$ is the curvature two-form given in Eq.(2.5), $\ast$ the Hodge duality transformation and $L_{2}$ the Weil polynomial formed by summation over algebraically independent elements of $S^{2}_{G}(\mathscr{G})$, will lead to the electroweak gauge field Lagrangian if $G=U(2)$, and to a gravitational gauge field Lagrangian if $G=SO(2,3)$.

The gauge invariance of (3.6), that is $L_{2}\textbf{(}\tilde{\Delta}(s^{\ast}(\tilde{\mu})),\ast\tilde{\Delta}(s^{\ast}(\tilde{\mu}))\textbf{)}=L_{2}\textbf{(}\tilde{\Delta}(\tilde{\mu}),\ast\tilde{\Delta}(\tilde{\mu})\textbf{)}$ for each diffeomorphism $s:P\rightarrow P$ such that $s(ug)=s(u)g$ and $\pi(u)=\pi(s(u)) $ for each $g\in G$ and $u\in P$ (Ref.13), follows from the $Ad(G)$-invariance of $L_{2}$ and the fact that $\tilde{\Delta}(\tilde{\mu})$ is a tensorial two-form on $P$.$^{[3,5]}$

In both cases, unitary and anti-de Sitter, the condition $[\gamma_{\ast}\mathscr{H},\mathscr{T}]\subset \mathscr{T}$ is satisfied as we will see in the next sections. Then $\Delta$, $\Omega$, $\Phi$, and $\Sigma$ as given in Eqs.(2.6)-(2.9) are all tensorial two-forms on $Q$, i.e.,
\begin{equation}
L_{2}(\Delta,\ast\Delta)=\gamma^{\ast}L_{2}(\tilde{\Delta},\ast\tilde{\Delta})
\end{equation}
projects to a unique four-form on $M$ such that
\begin{equation}
L_{2}(\Delta,\ast\Delta)=\pi^{\ast}\mathscr{L}.
\end{equation}

To determine  Weil Polynomials on an arbitrary Lie algebra $\mathscr{G}$, use can be made of the isomorphism between the algebra $S_{G}(\mathscr{G})$ of symmetric $Ad(G)$-invariant multilinear mappings on $\mathscr{G}$ and the algebra $P_{G}(\mathscr{G})$ of homogeneous $Ad(G)$-invariant polynomial functions on $\mathscr{G}$. The identifying isomorphism $\mathscr{S}:P_{G}(\mathscr{G})\rightarrow S_{G}(\mathscr{G})$ is determined by$^{[14]}$
\begin{equation}
(\mathscr{S}f)(X_1,...,X_k)=a_{i_{1}...i_{k}}\xi^{i_{1}}(X_1)\cdot\, \cdot\cdot\cdot\, \cdot\xi^{i_{k}}(X_k), \quad X_1,...,X_k \in\mathscr{G},
\end{equation}
where $a_{i_{1}...i_{k}}\xi^{i_{1}}\cdot\, \cdot\cdot\cdot\, \cdot\xi^{i_{k}}$ is the unique expression for the polynomial function of degree $k$ in the basis $\left\lbrace \xi^{i}\right\rbrace $ for $\mathscr{G}^{\ast}$ the dual space of $\mathscr{G}$.
  
Algebraically independent and generating $Ad(U(n))$-invariant polynomial functions on the Lie algebra $u(n)$ are given by the characteristic coefficients $f_k(X)$ in$^{[4]}$
\begin{equation}
det(\lambda I_n + iX)=\sum_{k=0}^{n}(-1)^k f_k (X)\lambda^{n-k},\; \textrm{for}\, X\in u(n).
\end{equation}
If $n=2$ and $\lbrace\chi^{i}\medskip_{j}\rbrace$ is a basis for $u(2)^{\ast}$ such that $\chi^{i}\medskip_{j}(X)=X^{i}\medskip_{j}$ for $X\in u(2)$, then the\\[-5mm] polynomial function of degree 2 is given by
\begin{equation}
f_{2}(X)=-\tfrac{1}{2}(X^{i}\medskip_{i}X^{j}\medskip_{j}-X^{i}\medskip_{j}X^{j}\medskip_{i}).
\end{equation}
The Weil polynomial corresponding to $f_{2}$ under the isomorphism $\mathscr{S}$ given in Eq.(3.9) is then
\begin{equation}
L_{2}(X_{1},X_{2})=-\tfrac{1}{2}(X^{i}_{1i}X^{j}_{2j}-X^{i}_{1j}X^{j}_{2i}),\;\textrm{for}\, X_{1},X_{2}\in u(2).
\end{equation}
So, all $Ad(U(2))$-invariant Weil polynomials on $u(2)$ of degree $2$ are polynomials proportional to the foregoing. In the basis
\begin{equation}
\lbrace T_{1},T_{2},T_{3},T_{4},\rbrace=\lbrace \tfrac{1}{2} i\tau_{1},\tfrac{1}{2} i\tau_{2},\tfrac{1}{2} i\tau_{3},\tfrac{1}{2} iI\rbrace,
\end{equation}
for $u(2)$, where $\tau_{1},\tau_{2},\tau_{3} $ are the Pauli matrices and $I$ the $2\times 2$ identity matrix, we find that these Weil polynomials are determined by
\begin{equation}
\begin{aligned}
&L_{2}(T_{a},T_{b})=C_{1}\delta_{ab}, \; a=1,2,3,\\
&L_{2}(T_{4},T_{4})=C_{2},
\end{aligned}
\end{equation}
where we introduced constants $C_{1}, C_{2}$ since the Weil polynomials are defined only upon an arbitrary constant.

For $G=SO(2,3)$ one can use the same methods to find that all $Ad(SO(2,3))$-invariant Weil polynomials on $so(2,3)$ of degree $2$ are polynomials proportional to 
\begin{equation}
L_{2}(X_{1},X_{2})=\tfrac{1}{2}(X^{A}_{1A}X^{B}_{2B}-X^{A}_{1B}X^{B}_{2A}),\; \textrm{for} \,X_{1},X_{2}\in so(2,3),
\end{equation}
with $X^{A}\medskip_{B}=\chi^{A}\medskip_{B}(X),\; \lbrace\chi^{A}\medskip_{B}\rbrace$ a basis for $so(2,3)^{\ast}$. The generators $J_{AB}$ of the Lie algebra\\ [-5mm] $so(2,3)$ are in $5\times 5$-matrix representation given by
\begin{equation}
(J_{AB})^{K}\medskip_{L}=\delta^{K}\medskip_{A}\eta_{BL}-\delta^{K}\medskip_{B}\eta_{AL}, \; A,B,K,L=0,1,2,3,4,
\end{equation}
\\[-10mm]where $\eta_{AB}=diag(-1,1,1,1,-1)$.  The ten elements $J_{AB}=-J_{BA}$ can be split into six generators $J_{ab}$, $a,b=0,1,2,3$
 of the $SO(1,3)$ subgroup and four anti-de Sitter boosts $P_{a}=(1/l)J_{4a}$, $l$ the de Sitter length. In the basis $\lbrace J_{ab},P_{c}\rbrace$ thus defined, we obtain from (3.15) and (3.16) that
\begin{equation}
\begin{aligned}
&L_{2}(J_{ab},J_{cd})=C_{1}(\eta_{ac}\eta_{bd}-\eta_{ad}\eta_{bc})=C_{1}\eta_{ab,cd},\\
&L_{2}(P_{a},P_{b})=C_{2}(1/l)^{2}\eta_{ab},\\
&L_{2}(J_{ab},P_{c})=0,
\end{aligned}
\end{equation}
where $C_{1}$ and $C_{2}$ are arbitrary constants.

\section{The electroweak gauge field Lagrangian}
In the electroweak theory, the photon field and massive
vector boson fields are constructed from the $SU(2)\times U(1)$
gauge fields and a suitable scalar Higgs field that reduces the
original symmetry to a $U(1)$ subsymmetry. The Lie algebra
$u(2)=su(2)\oplus u(1)$ of $SU(2)\times U(1)$ is, however, also the
Lie algebra of $U(2)$. A change from $SU(2)\times U(1)$ to $U(2)$
symmetry will therefore alter only few of the detailed mass
calculations in the model (see remark in Ref.11 ). Throughout the remainder of the section we then assume that (i)
$P(M,G)$ is a $U(2)$ principal fiber bundle over four-dimensional space-time M; (ii) $Q(M,H)$ is a $U(1)$ subbundle of $P$. More precisely, the subgroup $H$ is the charge conservation subgroup $U_{c}(1)$ of $U(2)$, i.e.,
\begin{equation}
A\in H \;\textrm{if and only if}\; A=\left( \begin{matrix}
 e^{2i\theta} & 0 \\ 
 0 & 1  
  \end{matrix}\right). 
\end{equation}

The structure group $U_{c}(1)$ of $Q$ is the isotropy subgroup of $U(2)$ at points $\smash{{0\choose a}\in \mathbb{C}^{2}}$, $a\in\mathbb{R}^{+}$ (Ref.11). The orbit space $\smash{V_{0}=U(2)\cdot {0\choose a}}$ is $\overline{S}$, the sphere of radius $a$ in $\mathbb{C}^{2}$, and is diffeomorphic with the coset manifold $U(2)/U_{c}(1)$. Therefore, the  existence of the reduced subbundle $Q$  of $P$ implies the existence of a global section in the vector bundle $E(M,\overline{S},U(2),P)$ associated with $P$, or equivalently of a symmetry breaking Higgs field $\psi :P\rightarrow \overline{S}$ such that $\psi^{-1}{0\choose a}=Q$.

In the basis (3.13), a connection one-form $\tilde{\mu}$ on $P$ may be written as
\begin{equation}
\tilde{\mu}=\tilde{a}^{1}gT_{1}+\tilde{a}^{2}gT_{2}+\tilde{a}^{3}gT_{3}+\tilde{a}^{4}g'T_{4},
\end{equation}
where $g$ and $g'$ are coupling constants. We then define the new basis $\lbrace\bar{T}_{1},\bar{T}_{2},\bar{T}_{3},\bar{T}_{4}\rbrace$, where
\begin{align}
&\bar{T}_{1}=(gT_{1}-ig T_{2})/\sqrt{2},\\
&\bar{T}_{2}=(gT_{1}+ig T_{2})/\sqrt{2},\\
&\bar{T}_{3}=-\cos \alpha (gT_{3})+\sin \alpha (g'T_{4}),\\
&\bar{T}_{4}=\sin \alpha (gT_{3})+\cos \alpha (g'T_{4}).
\end{align}
In this basis, the restriction $\mu$ of $\tilde{\mu}$ to $Q$ is written as
\begin{equation}
 \mu =W^{-}\bar{T}_{1}+W^{+}\bar{T}_{2}+Z\bar{T}_{3}+A\bar{T}_{4},
\end{equation} 
where
\begin{align}
&W^{-}=(a^{1}+ia^{2})/\sqrt{2},\\
&W^{+}=(a^{1}-ia^{2})/\sqrt{2},\\
&Z=-\cos \alpha\, a^{3}+\sin \alpha\, a^{4},\\
&A=\sin \alpha\, a^{3}+\cos \alpha\, a^{4}.
\end{align} 
It is clear that we identify $\alpha$ as the Weinberg angle $\theta_{w}$, such that the charge $e$ of the electromagnetic field A is given by
\begin{equation}
e=g\sin\alpha=g'\cos\alpha=gg'\left[g^{2}+(g')^{2}\right]^{-1/2}.
\end{equation}
Since $\bar{T}_{4}=e(T_{3}+T_{4})$ generates the Lie algebra $\mathscr{H}=u_{c}(1)$ of the charge conservation subgroup, and since the non-zero commutators of the new basis satisfy
\begin{align}
&\left[\bar{T}_{1},\bar{T}_{4}\right]=ie\bar{T}_{1},\\
&\left[\bar{T}_{2},\bar{T}_{4}\right]=-ie\bar{T}_{2},\\
&\left[\bar{T}_{1},\bar{T}_{2}\right]=ig(\cos\alpha\,\bar{T}_{3}-\sin\alpha\,\bar{T}_{4}),\\
&\left[\bar{T}_{1},\bar{T}_{3}\right]=-ig\cos\alpha\,\bar{T}_{1},\\
&\left[\bar{T}_{2},\bar{T}_{3}\right]=ig\cos\alpha\,\bar{T}_{2},
\end{align}
we can, according to Sec.2, decompose $\mu$ into two pieces:
\begin{equation}
\mu=\omega+\phi,
\end{equation}
where $\omega=A\bar{T}_{4}$ is a connection one-form on $Q$ corresponding to the electromagnetic gauge field $A$, and $\phi=W^{-}\bar{T}_{1}+W^{+}\bar{T}_{2}+Z\bar{T}_{3}$ is a tensorial one-form on $Q$ whose components correspond to the vector bosons $W^{+},\,W^{-}$, and $Z$ (see also Ref.11).

The reduction to $Q$ of the curvature obtained from a connection one-form on $P$ and given in Eqs. (2.6)-(2.9) is calculated making use of the commutation relations (4.13)-(4.17). We find that the components in $\Delta=\Omega+\Phi+\Sigma$ are
\begin{align}
&\Omega=(dA)\bar{T}_{4},\\
&\Phi=(DW^{-})\bar{T}_{1}+(DW^{+})\bar{T}_{2}+(dZ)\bar{T}_{3},\\
&\Sigma=-i\cos\theta_{w}(W^{-}\wedge Zg\bar{T}_{1}-W^{+}\wedge Zg\bar{T}_{2}-W^{-}\wedge W^{+}g\bar{T}_{3}+W^{-}\wedge W^{+}g'\bar{T}_{4}),
\end{align}
where
\begin{align}
&DW^{-}=dW^{-}-ieA\wedge W^{-},\\
&DW^{+}=dW^{+}+ieA\wedge W^{+}.
\end{align}
Equation (2.11) is now the Bianchi identity for the electro-magnetic field and (2.12) the corresponding identity for the intermediate vector boson field.

To obtain the electroweak gauge field Lagrangian we must substitute the explicit expressions for the components $\Omega,\,\Phi,\,\Sigma$ of $\Delta$ as given in Eqs.(4.19)-(4.23) into the Lagrangian (3.8) and apply Eq.(3.3). Then we need the expressions for $L_{2}(\bar{T}_{i},\bar{T}_{j})$, $i,j=1,2,2,4$, which one obtains from Eqs.(3.14) and the definitions (4.3)-(4.6) for the basis $\left\lbrace\bar{T}_{i}\right\rbrace$. To obtain dimensionless numbers for $L_{2}(\bar{T}_{i},\bar{T}_{j})$ we put $C_{1}=-1/(2g^{2})$ and $C_{2}=-1/(2g'^{2})$ in (3.14), which are also chosen such that in the final Lagrangian, terms appear with the correct coefficients. Then we find that
\begin{equation}
L_{2}(\bar{T}_{1},\bar{T}_{2})=L_{2}(\bar{T}_{3},\bar{T}_{3})=L_{2}(\bar{T}_{4},\bar{T}_{4})=-1/2
\end{equation}
and all other zero. For the Lagrangian we obtain
\begin{equation}
L_{2}(\Delta, \ast\Delta)=\mathscr{L}^{\gamma}+\mathscr{L}^{W}+\mathscr{L}^{Z}+\mathscr{L}^{W,W}+\mathscr{L}^{W,\gamma,Z},
\end{equation}
where
\begin{equation}
\mathscr{L}^{\gamma}=-\frac{1}{2}dA\wedge\ast(dA)
\end{equation}
is the Lagrangian of the electromagnetic field,
\begin{equation}
\mathscr{L}^{W}=-dW^{-}\wedge\ast (dW^{+})
\end{equation}
is the Lagrangian of the free charged vector boson field without the mass term,
\begin{equation}
\mathscr{L}^{Z}=-\frac{1}{2}dZ\wedge\ast(dZ)
\end{equation}
is the Lagrangian of the free neutral vector boson field without the mass term,
\begin{equation}
\mathscr{L}^{W,W}=\frac{1}{2}g^{2}W^{-}\wedge W^{+}\wedge \ast(W^{-}\wedge W^{+})
\end{equation}
is the self-interaction term of the $W$ bosons, and
\begin{multline}
\mathscr{L}^{W,\gamma,Z}=-\cos\theta_{w}\lbrace i(gdZ-g'dA)\wedge\ast(W^{-}\wedge W^{+})+i(gZ-g'A)\wedge\lbrack\ast(dW^{+})\wedge W^{-}\\
-\ast(dW^{-})\wedge W^{+}\rbrack +\cos\theta_{w}\lbrack gg'\textbf{(}A\wedge W^{-}\wedge\ast(W^{+}\wedge Z)+Z\wedge W^{-}\wedge\ast(W^{+}\wedge A)\textbf{)}\\
+g'^{2}A\wedge W^{-}\wedge\ast(A\wedge W^{+})+g^{2}Z\wedge W^{-}\wedge\ast(Z\wedge W^{+})\rbrack\rbrace
\end{multline}
is the interaction between the electromagnetic, $Z$, and $W$ boson fields.

In the electroweak theory, the gauge symmetry is broken spontaneously. Therefore, the gauge field Lagrangian $\mathscr{L}$ on $M$, of the type determined in (3.8), is supplemented with a $G$-invariant Lagrangian $\mathscr{L}^{\Psi}$ associated with a doublet of complex scalar physical  Higgs fields $\Psi :M\rightarrow \mathbb{C}^{2}\supset\bar{S}(a)$ with quartic self-interaction potential that assumes a minimum for $\Psi^{\dagger}\Psi=a^{2}$ [i.e., for  $\Psi (x)\in \bar{S}(a)$]. The effect of the gauge fields entering the covariant derivatives of the Higgs fields in $\mathscr{L}^{\Psi}$ is to give the $W$ and $Z$ fields a mass term such that $m_{W}^{2}=a^{2}g^{2}/2$ and $m_{Z}^{2}=a^{2}(g^{2}+g'^{2})/2$ (Ref.15).

\section{Gravitational gauge field Lagrangian}
The derivation of the gravitational gauge field Lagrangian is based on the anti-de Sitter group $G=SO (2,3)$ and its subgroup $H=SO (1,3 )$. Fundamental ingredients are (i) an anti-de Sitter principal bundle $P(M,G)$ over four-dimensional space-time M and a connection one-form $\tilde{\mu}$ on $P$; (ii) a Lorentz subbundle $Q(M,H)$ of $P$.

However, to interpret the resulting theory as a gravitational theory, the restricted curvature $\Delta=\gamma^{\ast}\tilde{\Delta}$ on $Q$ must be
directly related to the curvature of the underlying space-time manifold. Therefore, $Q(M,H)$ must be the bundle $O(M)$ of orthonormal frames over $M$, and $P(M,G)$ the anti-de Sitter frame bundle.

The coset space $G/H$ is now the anti-de Sitter space $ {F= SO(2,3)/SO(1,3)}$. It is a noncompact space of constant curvature $-1/l^{2}$ ($l$ the curvature radius) that can be represented by a hypersurface $H^{4}_{1}(l)=\lbrace\xi\in\mathbb{R}^{5}_{2};<\xi,\xi>=-l^{2}\rbrace$ in $\mathbb{R}^{5}_{2}$, $<,>$ being the quadratic form on $\mathbb{R}^{5}_{2}$ with signature $(-,+,+,+,-)$. The coset space $F$ can also be thought as the orbit of the $H$-fixed point $\xi_{0}=(0,0,0,0,l)\in \mathbb{R}^{5}_{2}$,  and $SO (1,3)$ is the isotropy subgroup of $SO (2,3)$ in $\xi_{0}$. The existence of the reduced subbundle $Q$ of $P$ now implies the existence of a global section in the with $P$ associated anti-de Sitter vector bundle $E(M,H^{4}_{1},SO(2,3),P)$ or of a symmetry breaking Higgs field $\psi:P\rightarrow H^{4}_{1}$ such that $\psi^{-1}(\xi_{0})=Q$.

The Lie algebra $\mathscr{G}=so(2,3)$ is defined by
\begin{align}
&\left[J_{ab},J_{cd}\right]=J_{ad}\eta_{bc}+J_{bc}\eta_{ad}-J_{ac}\eta_{bd}-J_{bd}\eta_{ac},\\
&\left[J_{ab},P_{c}\right]=P_{a}\eta_{bc}-P_{b}\eta_{ac},\\
&\left[P_{a},P_{b}\right]=(1/l^{2})J_{ab}.
\end{align} 
The parameter $l$ is still the radius of curvature of the anti-de
Sitter space whose group of isomorphisms is generated by this anti-de Sitter algebra. The generators $J_{ab}$ span the subalgebra $\mathscr{H}=so(1,3)$ of $\mathscr{G}$. The $P_{a}$, span a vector space $\mathscr{T}=\mathbb{R}^{4}_{1}$, such that Eq.(2.3) and the condition $\left[\gamma^{\ast}\mathscr{H},\mathscr{T}\right]\subset\mathscr{T}$ are satisfied. Therefore, if $\tilde{\mu}$ is a connection
one-form on $P$, its reduction $\mu$ to $Q$ splits as in Eq.(2.4).
Explicitly we may write
\begin{equation}
\mu=\omega +\phi=\frac{1}{2}\omega^{ab}J_{ab}+\phi^{a}P_{a},
\end{equation}
where now $\omega$ is $\mathscr{H}$-valued Lorentz connection while $\phi$ the tensorial one-form of type $(Ad H,\mathscr{T} )$ on $Q$ will be identified with the canonical form $\theta$ on $Q=O(M)$. Indeed, on $O(M)$ the canonical or soldering form $\theta$ has the same transformation low (2.1) as $\phi$ in (5.4). Therefore, it is possible to define a connection $\tilde{\mu}$ on $P$ for which $\phi=\theta$ (Ref.16). This identification implies that we take $\mu$ as the restriction to $Q$ of a
Cartan connection $\tilde{\mu}$ on $P$ and that the associated anti-de
Sitter vector bundle $E$ is a soldered bundle:$^{[17]}$ for all $x\in M$, the
fiber $F_{x}$ over $x$ is tangent to the base space $M$ at $\xi_{0}$, and the tangent spaces $T_{\xi_{0}}(F_{x})$ and $T_{x}(M)$ can be identified by an isomorphism. Moreover, the zero section $\xi (x)=\xi_{0}$ on $E$ can be identified with the base space $M$. After identification of $\phi$ with the canonical form $\theta$ on $Q$, $\Omega$ and $\Phi$ as given in Eqs.(2.7) and (2.8) can, respectively, be identified with the curvature form $\Omega$ and torsion form $\Theta$ of the bundle $O(M)$ . Explicitly. we can write
\begin{equation}
\Delta=\Omega+\Theta+\Sigma=\frac{1}{2}\left[\Omega^{ab}+(1/l^{2})\theta^{a}\wedge\theta^{b}\right]J_{ab}+\Theta^{c}P_{c},
\end{equation}
where
\begin{align}
&\Omega^{ab}=D\omega^{ab}=d\omega^{ab}+\omega^{ac}\wedge\omega^{b}\medskip_{c},\\
&\Theta^{a}=D\theta^{a}=d\theta^{a}+\omega^{ac}\wedge\theta_{c}.
\end{align}
From the Bianchi identities (2.11) and (2.12) we find
\begin{align}
&D\Omega^{ab}=d\Omega^{ab}+\omega^{ac}\wedge\Omega^{b}\medskip_{c}-\Omega^{ac}\wedge\omega^{b}\medskip_{c}=0,\\
&D\Theta^{a}=d\theta^{a}+\omega^{a}\medskip_{b}\wedge\Theta^{b}=\Omega^{ab}
\wedge\theta_{b}.
\end{align}

Substituting the expression (5.5) for $\Delta$ into the Lagrangian form (3.8) on $Q$, we find making use of (3.3) and Eqs.(3.17) that
\begin{multline}
  L_{2}(\Delta,\ast\Delta)=C_{1}(\frac{1}{2}\Omega^{ab}\wedge\ast\Omega_{ab}+\frac{1}{2l^{2}}\epsilon_{abcd}\Omega^{ab}\wedge\theta^{c}\wedge\theta^{d} \\
   +\frac{1}{4l^{4}}\epsilon_{abcd}\theta^{a}\wedge\theta^{b}\wedge\theta^{c}\wedge\theta^{d})+C_{2}\frac{1}{l^{2}}\Theta^{a}\wedge\ast\Theta_{a},  
   \end{multline} 
where $\epsilon_{0123}=1$. Except for the coefficients, this Lagrangian
is the same as the $SO(1,4)$ gauge field Lagrangian obtained by Townsend who used the diagonal $\eta_{ab}$ to contract group indices, or the Lagrangian of Zardecki who used the $SO(1,4)$ Cartan metric to contract the Yang-Mills indices.$^{[7]}$

The Lagrangian must be supplemented with an appropriate matter Lagrangian, representing the sources of the gravitational fields. Then the field equations are obtained by variation of this total Lagrangian with respect to $\theta^{a}$ and $\omega^{ab}$ separately. For the resulting field equations and their solutions we refer to section (5.2) of Ref.8. 

If the $SO(2,3)$ gauge theory is broken spontaneously, i.e., there exists a scalar field $\Psi:M\rightarrow\mathbb{R}^{5}_{2}\supset H^{4}_{1}(l)$ such that the potential $V(\Psi)$ in a gauge invariant Lagrangian $L^{\Psi}$, associated with $\Psi$, assumes a minimum for $<\Psi,\Psi>=-l^{2}$, that is for $\Psi(x)\in H^{4}_{1}(l)$, then the mass term given to the $\theta$ field and induced by the $SO(2,3)$ covariant derivatives of $\Psi$ will contribute to the vacuum energy density.
\section{Discussion}
In the derivation of the electroweak and gravitational gauge field Lagrangian in Secs.4 and 5 it has become clear that fiber bundle reduction and the closely related concept of (possibly spontaneous) symmetry breaking, described in Sec.2, are essential for both theories. Moreover, this geometrical framework makes it possible to describe both theories with one type of Lagrangian, Eq. (3.6). Fields in both theories of the same geometrical origin are the $W$ and $Z$ vector boson fields and the soldering form $\theta$ for the tensorial component of the reduced connection, and the electromagnetic field $A$ and Lorentz connection $\omega$ for the connection part. Other geometrical concepts that correspond are the soldering of the anti-de Sitter bundle, which is essential in the interpretation of the $SO(2,3)$ gauge theory as a gravitation theory, and the process of symmetry breaking in electroweak theory. In the electroweak theory the manifold of vacuum states is the sphere $\bar{S}(a)$ in $\mathbb{C}^{2}$ while the vacuum state corresponds to the global section of $E(M,\bar{S}(a), U(2),P)$, determined by $\smash{\Psi(x)={0\choose a}}$. In the $SO(2,3)$ gravitational gauge theory the role of the vacuum manifold is played by the anti-de Sitter space $H^{4}_{1}(l)\subset\mathbb{R}^{5}_{2}$ while the vacuum state corresponds to the global section of the anti-de Sitter bundle $E(M,H^{4}_{1}(l),SO(2,3),P)$ determined by $\Psi(x)=\xi_{0}=(0,l)$, which section can be identified with the space-time $M$ such that the anti-de Sitter bundle is "soldered". Then, in a spontaneously broken $SO(2,3)$ gauge theory, four-dimensional space-time could be interpreted as the "vacuum expectation value" of a physical Higgs field $\Psi:M\rightarrow\mathbb{R}^{5}_{2}$ that breaks $SO(2,3)$ symmetry down to the $SO(1,3)$ subsymmetry thereby creating what is called gravitational interaction.

\newpage 
\begin{center}
\textbf{\Large References}
\end{center}

\noindent $[1]$ A. Trautman, Rep. Math. Phys. 1, \textbf{29} (1970).\\
$[2]$ C. Nash and S. Sen, Topology and Geometry for Physicists (Academic,
London, 1983)\\
$[3]$ T. Verwimp. Prog. Theor. Phys. \textbf{80}, 330 (1988).\\
$[4]$ S. Kobayashi and K. Nomizu, Foundations of Differential Geometry
(Interscience, New York. 1963), Vols. 1 and 2.\\
$[5]$ K. Kakazu and S. Matsumoto, Prog. Theor. Phys. \textbf{78}, 166, 932 (1987); \textbf{79}, l431 (1988); \textbf{80}, 1109 (1988).\\
$[6]$ D. Lovelock. J. Math. Phys. \textbf{12}, 498 (1971).\\
$[7]$ P. K. Townsend Phys. Rev. D \textbf{15}, 2795 (1977); A. Zardecki, J. Math.Phys. \textbf{29}, 1661 (1988).\\
$[8]$ T. Verwimp. arXiv:1006.1614v4 [gr.qc] \\
$[9]$ R. Gilmore. Lie Groups, Lie algebras, and Some of Their Applications (Wiley-Interscience, New York, 1974).\\
$[10]$ A. Trautman, Czech. J. Phys. B \textbf{29}. 107 (1979).\\
$[11]$ R.O. Fulp and L. K. Norris, J. Math. Phys. \textbf{24}, 1871 (1983).\\
$[12]$ R. Cianci, J. Math. Phys. \textbf{22}, 2759 (1981).\\
$[13]$ D. Bleecker, Gauge Theory and Varuational Principles (Addison-Wesley, Reading, MA, 1981).\\
$[14]$ M. Spivak, A Comprehensive Introduction to Differential Geometry (Publish or Perish, Berkeley, CA, 1979), Vols. l-5.\\
$[15]$ J. Leite Lopes, Gauge Field Theories, An Introduction (Pergamon, New York (1981).\\
$[16]$ P. K. Smrz. J. Math. Phys. \textbf{28}, 2824 (l987).\\
$[17]$ W. Drechsler, J. Math. Phys. \textbf{18}, 1358 (1977).

\end{document}